\documentclass[%
preprint,
 amsmath,amssymb,
 aps,
prb,
]{revtex4-2}

\usepackage{graphicx}
\usepackage{dcolumn}
\usepackage{bm}
\usepackage{hyperref}
\usepackage[mathlines]{lineno}
\usepackage{xcolor}
\usepackage{adjustbox}

\begin{document}

\preprint{APS/123-QED}

\title{Spin-orbit torques and magnetotransport properties of $\alpha$-Sn and $\beta$-Sn heterostructures}



\author{Federico Binda}\affiliation{Department of Materials, ETH Zurich, CH-8093 Zurich, Switzerland.}
\author{Can Onur Avci}\affiliation{Department of Materials, ETH Zurich, CH-8093 Zurich, Switzerland.}
\author{Santos Francisco Alvarado}\affiliation{Department of Materials, ETH Zurich, CH-8093 Zurich, Switzerland.}
\author{Paul No\"el}\affiliation{Department of Materials, ETH Zurich, CH-8093 Zurich, Switzerland.}
\author{Charles-Henri Lambert}\affiliation{Department of Materials, ETH Zurich, CH-8093 Zurich, Switzerland.}
\author{Pietro Gambardella}
\affiliation{Department of Materials, ETH Zurich, CH-8093 Zurich, Switzerland.}
 
\email{federico.binda@mat.ethz.ch}



\date{\today}

\begin{abstract}
Topological insulators have emerged as an important material class for efficient spin-charge interconversion. Most topological insulators considered to date are binary or ternary compounds, with the exception of $\alpha$-Sn. Here we report a comprehensive characterization of the growth, magnetotransport properties, and current-induced spin-orbit torques of $\alpha$-Sn and $\beta$-Sn-based ferromagnetic heterostructures. We show that $\alpha$-Sn grown with a Bi surfactant on CdTe(001) promotes large spin-orbit torques in a ferromagnetic FeCo layer at room temperature, comparable to Pt, whereas $\alpha$-Sn grown without Bi surfactant and the non-topological phase, $\beta$-Sn, induce lower torques. The dampinglike and fieldlike spin-orbit torque efficiency in $\alpha$-Sn with Bi are 0.12 and 0.18, respectively. Further, we show that $\alpha$-Sn grown with and without Bi presents a spin Hall-like magnetoresistance comparable to that found in heavy metal/ferromagnet bilayers. Our work demonstrates direct and efficient charge-to-spin conversion in $\alpha$-Sn ferromagnetic heterostructures, showing that $\alpha$-Sn is a promising material for current-induced magnetization control in spintronic devices.

\end{abstract}

\maketitle


\section{Introduction}

Topological insulators (TIs) have attracted strong attention in spintronics as potential candidates for the efficient interconversion of charge and spin currents. TIs have a bulk band-gap and conducting states at their boundaries, which exhibit spin-momentum locking \cite{G203, G204, G4, G136}. Owing to this feature, electric conduction via the TIs' surface states can be harnessed for the generation of spin currents in spintronic devices. There exist already numerous studies where TIs have been used for generating spin-orbit torques (SOTs) in ferromagnetic heterostructures \cite{G121,G123, G124, G128, G126, G197, G206, G205}. Finding new TIs and characterizing the existing ones constitutes an important challenge of spintronics research.

SOTs are current-induced magnetic torques capable of manipulating the magnetization of magnetic thin films and nanostructures. SOTs were originally discovered and extensively studied in heavy metal/ferromagnet bilayers \cite{ando2008electric, G118, miron2011perpendicular, liu2011spin, G19, G117, G50, kim2013layer, liu2012spin, manchon2009theory} and noncentrosymmetric magnetic semiconductors \cite{chernyshov2009evidence, kurebayashi2014antidamping}. In these systems, SOTs rely on charge-to-spin conversion due to interfacial and bulk mechanisms, such as the Rashba-Edelstein and spin Hall effect, respectively \cite{G117}. 

Since the first reports of SOTs from TIs \cite{G121,G123}, extensive efforts have been made to explore TIs as efficient SOT generators. The origins of SOTs in the TI-based systems, however, are not yet well-understood. SOTs were initially attributed to the TIs' spin-momentum locked surface states \cite{G123}; this view is now under debate due to the possible, perhaps dominant, role of the bulk conduction in intrinsically doped TIs. The latter enables the bulk spin-Hall effect to become a source of SOTs \cite{G197}. Additionally, the formation of Rashba-split surface bands and the Rashba-Edelstein effect play an important role at TI/ferromagnet (FM) interfaces \cite{G124}. Therefore, disentangling the surface state contributions vs other potential SOT sources is not trivial. Moreover, it is unclear whether the properties of the TIs are preserved when they come in contact with magnetic materials, which is usually the case in SOTs studies. For instance, magnetic impurities may destroy the topological surface states \cite{zhang2016band}. Also, intermixing, commonly occurring at the interfaces between TIs and FMs, can be detrimental for SOTs \cite{G208, G205}. 

Whereas most SOT investigations have focused on Bi-based binary and ternary compounds such as Bi\textsubscript{2}Se\textsubscript{3}, Bi\textsubscript{x}Sb\textsubscript{1-x} and (Bi\textsubscript{x}Sb\textsubscript{1-x})\textsubscript{2}Te\textsubscript{3} \cite{G106, G99}, other TI systems such as $\alpha$-Sn have been much less studied despite their simpler stoichiometry \cite{G27, G29, G116, G207}. $\alpha$-Sn is an allotropic phase of Sn, which crystallizes in the diamond cubic structure. In the bulk form, $\alpha$-Sn is a zero-gap semimetal in which the usual conduction and valence bands are inverted. In the thin film form, strain and quantum confinement open a band-gap, making $\alpha$-Sn a topological insulator \cite{G27, G52, G29}. The presence of spin-polarized topological surface states has been confirmed by angle-resolved photoemission spectroscopy \cite{G27, G29, G207}. Recently, Rojas-Sánchez \textit{et al.} performed spin-pumping experiments on an $\alpha$-Sn/Ag/Fe heterostructures and reported a spin-to-charge conversion efficiency significantly larger than that of the 5$d$ heavy metals \cite{G31}. This finding makes $\alpha$-Sn a promising candidate for the reciprocal process of charge-to-spin conversion and efficient generation of SOTs. To our knowledge, measurements of SOTs  in $\alpha$-Sn-based magnetic heterostructures have not been reported so far.

Motivated by the spin-pumping experiments on $\alpha$-Sn \cite{G31} and the lack of SOT studies in this material, we conducted a comprehensive investigation of the growth, magnetotransport properties, and SOTs in a variety of Sn-based ferromagnetic heterostructures. The aim of this study is two-fold. First, we characterize the direct charge-to-spin conversion in $\alpha$-Sn by performing SOTs measurements. We compare the results obtained on FeCo deposited on epitaxial $\alpha$-Sn grown on CdTe(001) with and without Bi surfactant as well as on $\beta$-Sn grown on MgO(001)/FeCo. Second, we measure the magnetoresistance and anomalous Hall effect in these three systems, discussing their behavior in relationship to the SOTs. We find that $\alpha$-Sn with Bi surfactant presents the largest SOT efficiency, comparable to that of the 5$d$ heavy metals, whereas $\alpha$-Sn without Bi and $\beta$-Sn have efficiencies that are about 2 times and 4 times smaller, respectively. The fieldlike SOT is larger than the dampinglike SOT in all the heterostructures. The $\alpha$-Sn heterostructures also present a spin Hall-like magnetoresistance, which is another signature of strong charge-to-spin conversion in these systems. These findings show that $\alpha$-Sn is an efficient material for generating current-induced SOTs and highlight the importance of Bi for improving the SOT efficiency and modifying the magnetotransport properties of $\alpha$-Sn-based ferromagnetic heterostructures. 

This article is organized as follows. Section II describes the sample growth and structural characterization. In Sect. III we present the magnetic and electrical characterization of the Sn/FeCo heterostructures. In Sect. IV we present the SOTs measurement and compare the torque efficiencies. The results are discussed in Sect. V, followed by the conclusions given in Sect. VI.

\begin{figure*}
\includegraphics[scale=0.5]{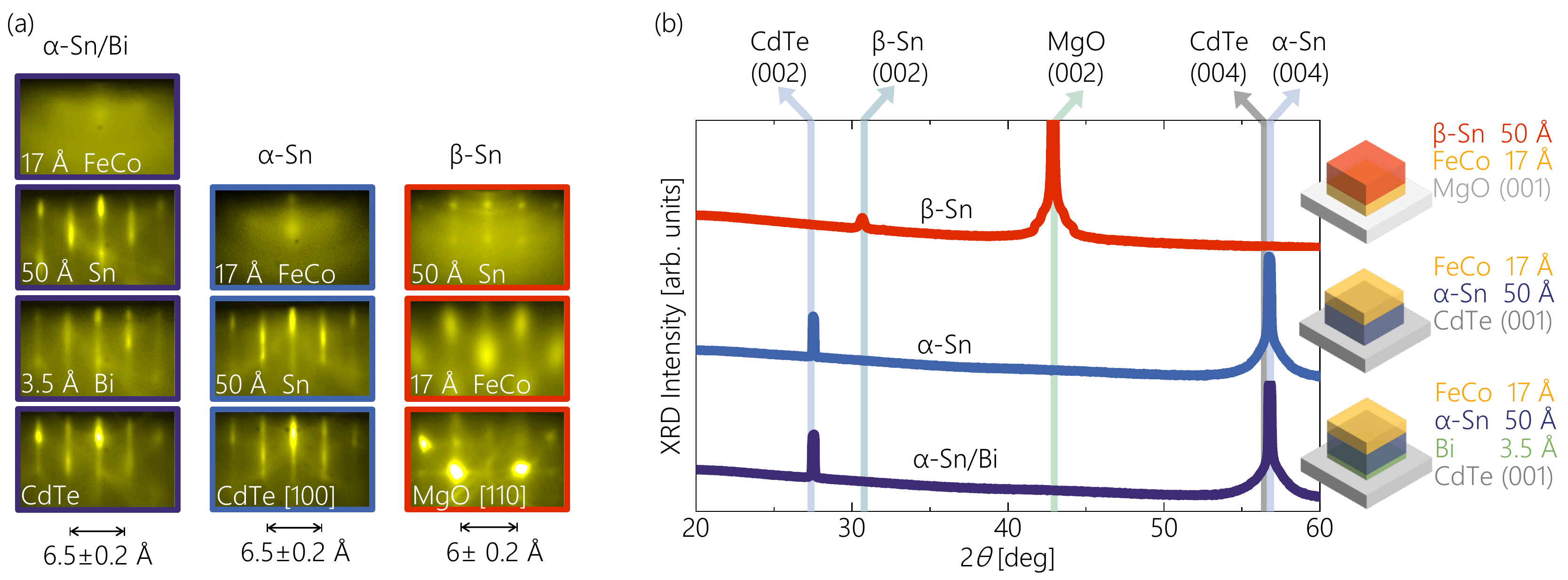}%
\caption{\label{fig:FIGURE1} (a) RHEED images of the different layers in the $\alpha$-Sn/Bi, $\alpha$-Sn and $\beta$-Sn heterostructures, each column corresponding to a sample and each row corresponding to a layer, from the substrates (bottom) to the last deposited layer (top) before capping. The RHEED patterns are along [100] for CdTe(001) and [110] for MgO(001). The lattice spacing associated to the distance in reciprocal space between relevant reflexes of the RHEED patterns are shown at the bottom of the figure. The sample with $\alpha$-Sn/Bi (left) and $\alpha$-Sn (center) show the RHEED patterns expected for epitaxial $\alpha$-Sn on CdTe. The sample with $\beta$-Sn (right) shows RHEED patterns of epitaxial FeCo and $\beta$-Sn. (b) X-ray diffractograms with a sketch of the corresponding heterostructures. $\alpha$-Sn/Bi and $\alpha$-Sn (bottom and center) show only the reflexes of the CdTe substrate, whereas $\beta$-Sn (top) present the reflexes of both epitaxial $\beta$-Sn and MgO substrate.}
\end{figure*}

\section{Samples}

\subsection{Substrate preparation and film growth}

The Sn and FeCo films were deposited at room temperature by molecular beam epitaxy in an ultra-high-vacuum system with base pressure $<$10\textsuperscript{-10}~mbar. High purity elements ($\geq$ 99.999$\%$) were evaporated from Knudsen cells with the following growth rates monitored by a quartz crystal balance: Bi (0.8~\AA/min), Sn (1.2~\AA/min), Fe (0.4~\AA/min), Co (0.37~\AA/min), Al (0.9~\AA/min). Fe and Co were co-deposited and Al was deposited with an O\textsubscript{2} pressure of 1.6$\times$10\textsuperscript{-6}~mbar. As mentioned earlier, we fabricated and studied three different heterostructures: (\mbox{\textit{i}) CdTe(001)/Bi[3.5~\AA]/$\alpha$-Sn[50~\AA]/FeCo[17~\AA]/AlO\textsubscript{x}}; (\mbox{\textit{ii}) CdTe(001)/$\alpha$-Sn[50~\AA]/FeCo[17~\AA]/AlO\textsubscript{x}};   (\mbox{\textit{iii}) MgO(001)/FeCo [17~\AA]/$\beta$-Sn[50~\AA]/AlO\textsubscript{x}}. Henceforth, we refer to those heterostructures as $\alpha$-Sn/Bi, $\alpha$-Sn, and $\beta$-Sn, respectively. Additionally, to serve as a reference, we deposited single films of $\alpha$-Sn[50~\AA] and FeCo[17~\AA] on CdTe(001) and a single film of FeCo[17~\AA] on MgO(001), all capped by AlO$\textsubscript{x}$.

Bulk $\alpha$-Sn is structurally stable only below 13.2~$^{\circ}$C, whereas thin films are stable above 100~$^{\circ}$C \cite{G75}. For this study, we selected a film thickness of 50~\AA, which is in the 39-55~\AA \space range in which strain and quantum confinement effects make $\alpha$-Sn a TI \cite {G27}. The epitaxial growth of $\alpha$-Sn (lattice constant $a=6.489~\rm \AA$) \space has been reported on InSb ($a=6.479~\rm \AA$) and CdTe ($a=6.482~\rm \AA$) \cite{G43, G71, G75}. Here we chose CdTe due to its higher resistivity with respect to InSb to minimize current shunting through the substrate in the electrical and SOT measurements.

The preparation of the CdTe substrates before the film deposition was crucial to grow high-quality $\alpha$-Sn films. We first sonicated the substrates in acetone, rinsed them in isopropanol and methanol, etched them for 20~s in a 0.5$\%$ bromine-methanol solution to remove the top pristine oxide layer, and finally rinsed them for 1~min in water to form a heat-sensitive TeO\textsubscript{x} film. Subsequently, the substrates were inserted into the vacuum chamber, sputtered with Ar-ions for 10~min, and heated to about 230$^\circ$C for 10~min to remove the TeO\textsubscript{x} film. The heterostructures were then obtained by subsequent deposition of the layers described above. In the $\alpha$-Sn/Bi sample, Bi was used because of its surfactant properties in the growth of Sn and Sn-based thin films: Bi promotes the growth of $\alpha$-Sn on InSb \cite{G27} and of SnGe on Ge \cite{lyman1996surfactant}. In both cases, Bi segregates to the surface of the growing film, which we expect to occur in our films as well. The preparation of the MgO substrate involved sonication in acetone, isopropanol, and methanol. The MgO substrate was then inserted into the vacuum chamber and heated to about 500$^\circ$C for 90~min. The deposition of the $\beta$-Sn heterostructure followed.

\subsection{Structural characterization}

The crystal structure of the heterostructures was characterized using a combination of \textit{in-situ} reflection high-energy electron diffraction (RHEED) and \textit{ex-situ} X-ray diffraction (XRD). RHEED was performed along the [100] direction for CdTe and the [110] direction for MgO. RHEED patterns were recorded for the prepared substrates and after every deposited layer, except for the  AlO\textsubscript{x} capping. The XRD of the full heterostructure was performed using a PANalytical X’Pert$\textsuperscript{3}$ diffractometer with Cu K-$\alpha$ radiation ($\lambda=1.5406~\rm \AA$). 

Figure~\ref{fig:FIGURE1}(a) shows the RHEED patterns from the three samples studied here. The two $\alpha$-Sn samples exhibit similar patterns for the CdTe substrates and $\alpha$-Sn films. These layers have thus similar lateral lattice spacing (indicated at the bottom), independently of the presence of the Bi surfactant layer. \color{black} The resemblance of the patterns proves the epitaxial growth of $\alpha$-Sn on CdTe(001). The streaky reflections, which are indicative of the two-dimensional character of the films, also present intensity spots that we attribute to the roughness of the underlying substrate.  \color{black} The RHEED pattern of the FeCo films grown on $\alpha$-Sn shows rings on a cloudy background, suggesting that the FeCo layers consist of both polycrystalline and amorphous phases. In the $\beta$-Sn sample, the RHEED patterns are similar for the MgO substrate and the FeCo layer, proving the epitaxial deposition of the latter. The $\beta$-Sn film presents a spotty pattern, indicating that the film consists of three-dimensional islands. The spacing of this pattern is nevertheless comparable with the base lattice constant of $\beta$-Sn ($a=b=5.832~\rm \AA$, $c=3.182~\rm \AA$, tetragonal structure).

Figure~\ref{fig:FIGURE1}(b) shows the XRD measurements of the three samples. The vertical lines indicate the 2$\theta$ angle of the relevant crystallographic reflexes. The (004) reflexes of CdTe and $\alpha$-Sn are indistinguishable, owing to their similar lattice constants. The XRD of the two $\alpha$-Sn samples shows only the reflexes of CdTe, as the FeCo layer is either too thin or  too disordered to be detected. The XRD measurement of the $\beta$-Sn sample shows the reflexes of MgO and epitaxial $\beta$-Sn, whereas the FeCo could not be detected. The RHEED and XRD measurements collectively indicate the successful deposition of $\alpha$-Sn and $\beta$-Sn films, as anticipated.
\color{black}
To shed light on the position of the Bi atoms in samples including $\alpha$-Sn we performed angle-dependent X-ray photoelectron spectroscopy (XPS) on a similar heterostructure consisting of \mbox{CdTe/Bi[3.5~\AA]/$\alpha$-Sn[50~\AA]/Co[17~\AA]/Al[15~\AA]}. The results of this study indicate that the largest concentration of Bi is located below the top Al layer. This confirms that Bi has a strong tendency to behave as a surfactant for Sn grown on CdTe and additionally diffuses and segregates through the magnetic layer. From the XPS data, however, we cannot exclude that some Bi atoms might be incorporated in the Sn and magnetic layer.
\color{black}

\section{Magnetic and electrical characterization}

\subsection{Magnetic properties}

\begin{figure*}
\includegraphics[scale=0.5]{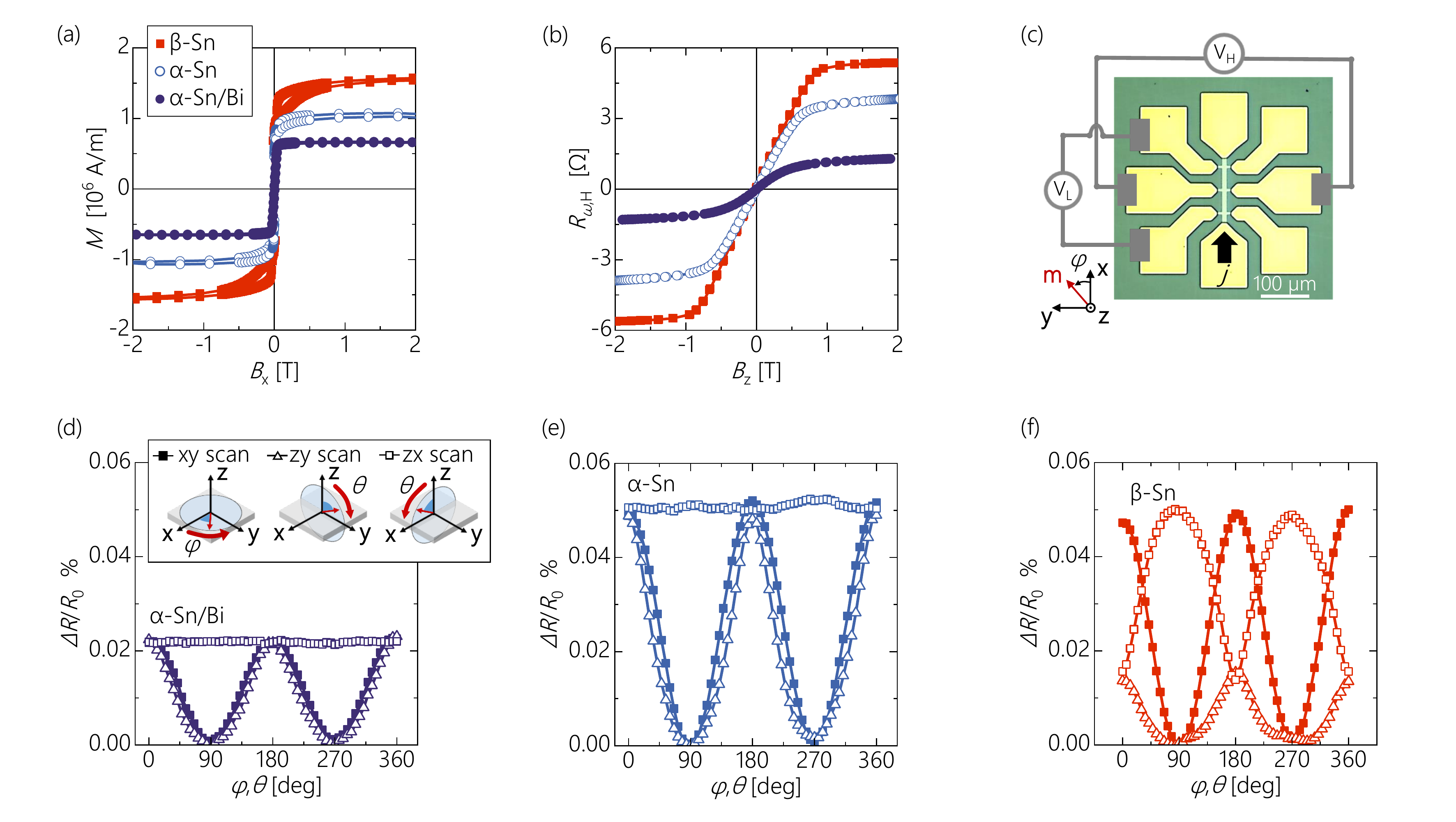}
\caption{\label{fig:FIGURE2} (a) Magnetic hysteresis loops as a function of in-plane magnetic field measured by SQUID. (b) Anomalous Hall resistance as a function of out-of-plane magnetic field. (c) Patterned device and coordinate system. The angle $\theta$ (not shown) is the polar angle between the z-axis and the magnetization vector $\textbf{m}$. (d, e, f)  Percent variation of the longitudinal resistance normalized with respect to $R_{\rm 0}=R(\textbf{m} \Vert \textbf{x})$ for the angular scans xy, zy and zx (see the inset). The scans are performed in a constant magnetic field $B_{\rm ext}=2.2$~T. Data are shown for $\alpha$-Sn/Bi(d), $\alpha$-Sn(e) and $\beta$-Sn (f). The current density used to perform the measurements shown in (b,d-f) is $j$=1.4$\times$10\textsuperscript{6}~A/cm\textsuperscript{2}. All measurements were performed at room temperature.}
\end{figure*}

We measured the magnetization of the as-grown layers as a function of in-plane magnetic fields using a superconducting quantum interference device (SQUID) at room temperature [see Fig.\ref{fig:FIGURE2}(a)]. The films have saturation magnetization $M$\textsubscript{s} = 0.7$\times$10\textsuperscript{6}~A/m ($\alpha$-Sn/Bi, solid blue circles), 1.0$\times$10\textsuperscript{6}~A/m ($\alpha$-Sn, open blue circles), and 1.5$\times$10\textsuperscript{6}~A/m ($\beta$-Sn, solid red squares). FeCo thin films grown on $\alpha$-Sn have a strongly reduced magnetization compared to those grown on MgO(001). \color{black} We attribute the larger $M$\textsubscript{s} of $\beta$-Sn with respect to the $\alpha$-Sn samples to the larger $M$\textsubscript{s} of the body-centered cubic (bcc) phase of FeCo  when grown epitaxially on MgO ($M$\textsubscript{s}$\approx$2$\times$10\textsuperscript{6}~A/m) \cite{G183}. We assume that FeCo grows face-centered cubic (fcc) on $\alpha$-Sn. This phase  has a lower $M$\textsubscript{s} compared to the bcc phase, as shown for FeCo grown on diamond ($M$\textsubscript{s}$\approx$1.4$\times$10\textsuperscript{6}~A/m) \cite{G202}, which has the same crystal structure as $\alpha$-Sn. \color{black} Moreover, issues common to ultrathin films such as interfacial mixing with nonmagnetic elements and dead layer formation lead to a further reduction of $M$\textsubscript{s} in our films, because the $M$\textsubscript{s} of our samples remains lower than either the bcc or fcc phase of bulk FeCo. The as-grown FeCo layers deposited on $\alpha$-Sn also possess a weak uniaxial magnetic anisotropy in the range of 1~mT along one of the $\langle110\rangle$ directions of CdTe(001). This phenomenon, typically observed in magnetic thin films grown on substrates with the zinc blende structure \cite{krebs1987properties},  indicates that the [1-10] and [110] directions of the $\alpha$-Sn(100) layer grown on CdTe(001) are not symmetrically equivalent, just as is the case for the substrate.

The magnetotransport measurements were performed on Hall bar devices of 10~$\mu$m width and 100~$\mu$m length fabricated by photolithography and ion milling [see Fig.~\ref{fig:FIGURE2}(c)]. The devices were wire bonded and mounted on a motorized stage allowing for in-plane and out-of-plane rotations in an electromagnet producing fields of up to 2.2~T at room temperature. Using the reference system shown in Fig.~\ref{fig:FIGURE2}(c), we define $\varphi$ as the azimuthal angle between the x-axis and the magnetization vector $\textbf{m}$, and $\theta$ as the polar angle between the z-axis and $\textbf{m}$. The measurements were performed using an a.c. current and recording the 1\textsuperscript{st} and 2\textsuperscript{nd} harmonic of the transverse and longitudinal resistances as a function of the magnetic field direction (angle scan) or amplitude (field sweep). All measurements were performed in the dark to suppress the photovoltaic effect of CdTe. We used current densities in the range of $j$=(1.4-2.8)$\times$10\textsuperscript{6}~A/cm\textsuperscript{2} with a frequency of $\omega$/2$\pi$=10~Hz. The current density $j$ is calculated by considering the thickness of the entire stack, excluding the substrate and the AlO\textsubscript{x} capping.

We measured the 1\textsuperscript{st} harmonic Hall signal ($R$\textsubscript{$\omega$,H}) as a function of out-of-plane magnetic field ($B$\textsubscript{z}) for characterizing the anomalous Hall resistance ($R$\textsubscript{AHE}) and the saturation field $B$\textsubscript{sat}, as shown in Fig.~\ref{fig:FIGURE2}(b). In this configuration, $R$\textsubscript{$\omega$,H} is proportional to the z-component of $\textbf{m}$. Because of the in-plane anisotropy, the field required to saturate the magnetization out-of-plane is defined as $B_{\rm sat}=B_{\rm dem}-B_{\rm PMA}$. Here $B_{\rm dem}$ is the demagnetizing field,  with $B_{\rm dem}$$\approx$$\mu_0M\textsubscript{s}$, and $B_{\rm PMA}$ is the effective perpendicular anisotropy field. \color{black} We then calculated $B_{\rm PMA}$, taking $M\textsubscript{s}$ from the SQUID measurements and $B_{\rm sat}$ from the out-of-plane field scan, and estimated the perpendicular magnetic anisotropy constant as $K_{\rm \perp}=(M_{\rm s}B_{\rm PMA})/2$  \cite{krishnan2016fundamentals}. \color{black} For the $\alpha$-Sn/Bi, $\alpha$-Sn and $\beta$-Sn samples, we found $B$\textsubscript{sat}=470, 625 and 780~mT, corresponding to $K_{\perp}$=1.4$\times$10\textsuperscript{5}, 3.2$\times$10\textsuperscript{5} and 8.3$\times$10\textsuperscript{5}~J/m$\textsuperscript{3}$, respectively. The difference in  $K$\textsubscript{$\perp$} values can be ascribed to the different interface contribution to the magnetic anisotropy. In particular for the $\beta$-Sn sample with FeCo epitaxially grown on MgO, we found an enhanced $K_{\perp}$. We also extracted $R$\textsubscript{AHE} = \mbox{[$R$\textsubscript{AHE}(\textbf{m}$\Vert \textbf{z})$\text{--}$R$\textsubscript{AHE}($\textbf{m} \Vert$\text{--}\textbf{z})]/2}, which gives $R$\textsubscript{AHE}=1.1~$\Omega$ for $\alpha$-Sn/Bi, 3.5~$\Omega$ for $\alpha$-Sn and 5.4~$\Omega$ for $\beta$-Sn. The factor three difference in $R$\textsubscript{AHE} between the two $\alpha$-Sn samples will be discussed in the context of SOTs in Sect. V.

\subsection{Magnetoresistance}
We further characterized the resistivity and magnetoresistance by measuring the 1$\textsuperscript{st}$ harmonic longitudinal resistance ($R_{\omega, \rm L}$). For $\alpha$-Sn/Bi, $\alpha$-Sn and $\beta$-Sn, the resistivities are $\rho$=179~$\mu\Omega$cm, 204~$\mu\Omega$cm, 342~$\mu\Omega$cm. The magnetoresistance $R_{\omega, \rm L}$ can be written as a function of the magnetization angles $\varphi$ and $\theta$ in its most general form as \cite{G41}:
\begin{equation}
R_{\omega,{\rm L}}=R_{\rm 0}-\Delta R_{\rm zx}~{\rm sin}^{2}\theta~{\rm cos}^{2}\varphi-\Delta R_{\rm zy}~{\rm sin}^{2}\theta~{\rm sin}^{2}\varphi
\label{eq:zero},
\end{equation}
where $R_{\rm 0}=R(\textbf{m} \Vert \textbf{x})$, $\Delta R_{\rm zx}$ is the resistance difference between magnetization pointing along the z-axis and the x-axis, $\Delta R_{\rm zy}$ is the resistance difference between magnetization pointing along the z-axis and y-axis, and $\Delta R_{\rm xy}=\Delta R_{\rm zy}-\Delta R_{\rm zx}$. We measured the magnetoresistance by rotating the devices in the xy, zy, zx planes in a static magnetic field of 2.2~T, which is large enough to saturate $\textbf{m}$ along any direction in all the samples. Figure~\ref{fig:FIGURE2}(d, e, f) shows the percent variation of $R_{\omega,{\rm L}}$ with respect to $R_{\rm 0}=R(\textbf{m} \Vert \textbf{x})$. Both the $\alpha$-Sn samples present similar magnetoresistance characterized by $\Delta R_{\rm xy}\approx\Delta R_{\rm zy}>0$ and $\Delta R_{\rm zx}\approx0$, with $\frac{\Delta R_{\rm xy}}{R_{\rm 0}}\approx$0.02 \% for $\alpha$-Sn/Bi and 0.05 \% for $\alpha$-Sn. This behavior differs from the typical anisotropic magnetoresistance of polycrystalline thin films, for which the resistance is larger when $\textbf{m}$ is orthogonal to the current, resulting in $\Delta R_{\rm xy}\approx\Delta R_{\rm zx}>0$ and $\Delta R_{\rm zy}\approx0$. According to observations made in in ultrathin magnetic layers in contact with heavy metals, different explanations have been proposed for this unusual magnetoresistance, which rely on the so-called spin Hall magnetoresistance (SMR) \cite{nakayama2013spin} or the anisotropic interface magnetoresistance (AIMR) \cite{kobs2011anisotropic}. In the first scenario, a large magnetoresistance appears in $\Delta R_{\rm zy}$ and $\Delta R_{\rm xy}$ due to the asymmetry in the absorption and reflection of the spin current generated by the bulk spin Hall effect of the nonmagnetic conductor upon the rotation of $\textbf{m}$ in the $zy$ and $xy$ planes \cite{lu2013hybrid, G41, kim2016spin}. In the second scenario, spin scattering due to a localized spin-orbit potential at the interface leads to a strong dependence of the magnetoresistance on the out-of-plane component of $\textbf{m}$, giving rise to a large $\Delta R_{\rm zy}$ \cite{grigoryan2014intrinsic, zhang2015anisotropic}. These effects are hard to separate in practice, but are both indicative of charge-spin conversion effects taking place in bilayer systems \cite{G182, G49}. Thus, the finite $\Delta R_{\rm zy}$ is a first signature that current injection in $\alpha$-Sn/FeCo layers can generate significant SOTs. In contrast, the $\beta$-Sn sample shows non-zero values for all the three magnetoresistances $\frac{\Delta R_{\rm xy}}{R_{\rm 0}}$, $\frac{\Delta R_{\rm zy}}{R_{\rm 0}}$ and $\frac{\Delta R_{\rm zx}}{R_{\rm 0}}$. We shall discuss these magnetoresistance data in the context of SOTs in Sect. V.

\section{Spin-orbit torques}

\subsection{Harmonic Hall voltage detection of spin-orbit torques}

\begin{figure}
\includegraphics[scale=0.5]{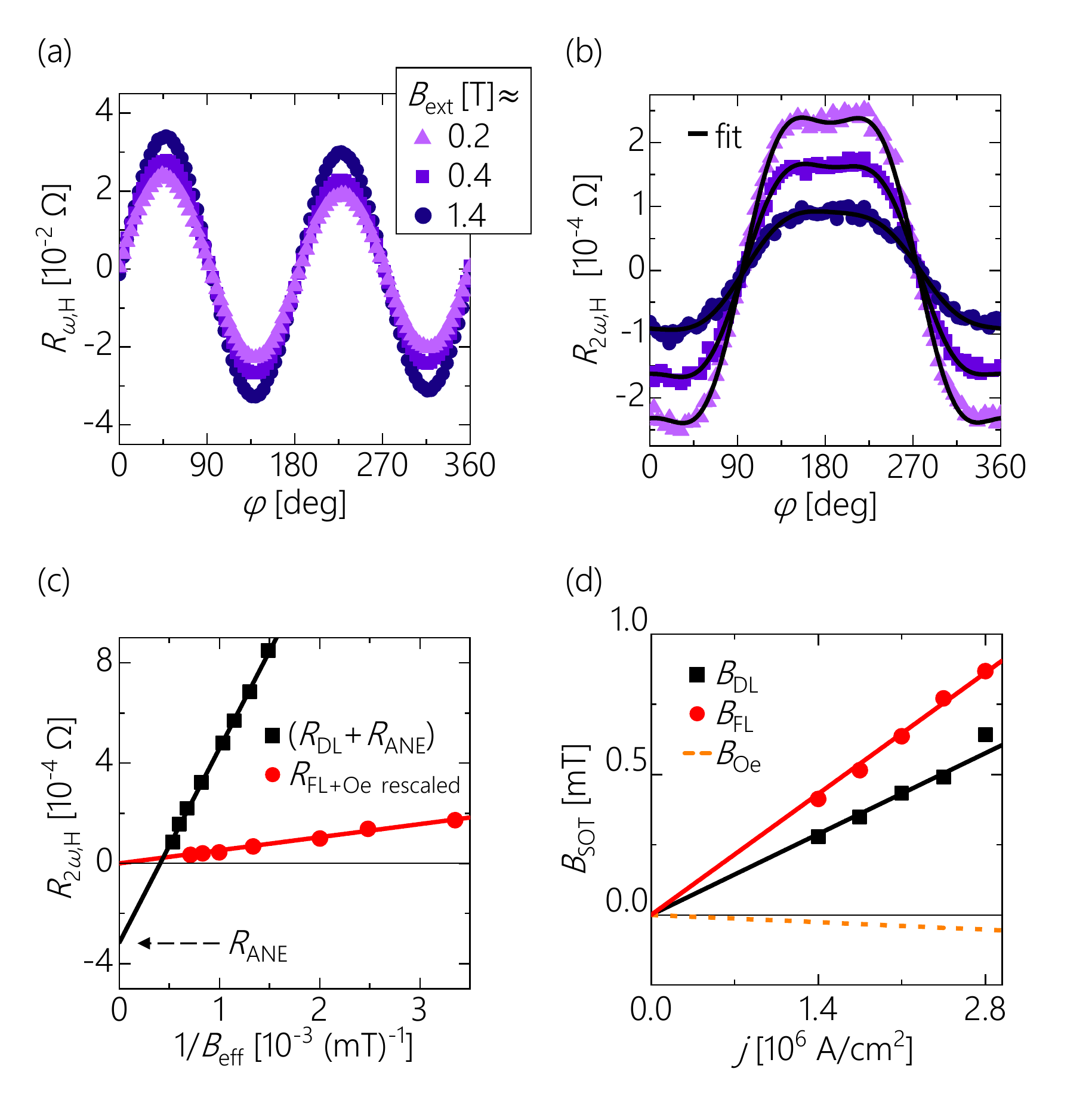}
\caption{\label{fig:FIGURE3} SOT quantification in $\alpha$-Sn/Bi. (a) Angle scan of $R_{\omega,{\rm H}}$ as function of $\varphi$ at different $B_{\rm ext}$. The oscillations of $R_{\omega,{\rm H}}$ are due to the planar Hall effect. (b) Angle scan of $R_{2\omega,{\rm H}}$ recorded simultaneously with $R_{\omega,{\rm H}}$ showing contributions from SOT, Oersted field and magnetothermal effects. Points are experimental data, lines are fits to Eq.~\ref{eq:two}. The current density in (a) and (b) is $j$=2.8$\times$10$\textsuperscript{6}$~A/cm$\textsuperscript{2}$. (c) Dependence of the FL-SOT resistance and DL-SOT resistance as a function of the inverse of their corresponding ${B_{\rm eff}}$ (defined in the main text). (d) Dependence of the SOT effective fields as function of the applied current density. The lines are linear fits to the data. The estimated Oersted field is plotted as a dashed line.}
\end{figure}

SOTs consist of two components: a fieldlike torque (FL-SOT, including the Oersted field contribution due to current flowing in the nonmagnetic layer)  \mbox{$\textbf{T}\textsubscript{FL+Oe}\sim\textbf{m}\times\textbf{y}$}, and a dampinglike torque (DL-SOT) \mbox{$\textbf{T}\textsubscript{DL}\sim\textbf{m}\times(\textbf{m}\times\textbf{y})$}. Their action on the magnetization can be described by effective fields that are accessible experimentally: \mbox{$\textbf{B}\textsubscript{FL+Oe}\sim\textbf{y}$} and \mbox{$\textbf{B}\textsubscript{DL}\sim\textbf{m}\times\textbf{y}$}. We measured the SOT effective fields 
of the three samples using the Harmonic Hall Voltage detection technique discussed in detail in Ref.~\onlinecite{G50}. The method involves recording the 1$\textsuperscript{st}$ and the 2$\textsuperscript{nd}$ harmonic Hall resistance at frequencies $\omega$ and 2$\omega$, respectively, as a function of the direction or intensity of an external magnetic field. Below, we present an exemplary step-by-step process explaining how we quantified the SOTs in the $\alpha$-Sn/Bi  sample. 

The 1$\textsuperscript{st}$ harmonic Hall resistance consists of contributions from the anomalous Hall ($R$\textsubscript{AHE}) and planar Hall effect ($R$\textsubscript{PHE}) that can be expressed as a function of the magnetization angles $\theta$ and $\varphi$:
\begin{equation}
R_{\omega,{\rm H}}=~R_{\rm AHE}~{\rm cos}~{\theta}~+~R_{\rm PHE}~{\rm sin^2}~{\theta}~{\rm sin}~{2\varphi}
\label{eq:one}.
\end{equation}
The 2$\textsuperscript{nd}$ harmonic Hall resistance consists of the resistance contributions from SOTs ($R$\textsubscript{FL+Oe} and $R$\textsubscript{DL}), which are due to the SOT-induced oscillations of the magnetization, and the magnetothermal effects ($R$\textsubscript{ANE}) due to the anomalous Nernst effect \cite{G19, G50}. When the magnetization lies in-plane, as it is the case for our samples, the 2$\textsuperscript{nd}$ harmonic Hall resistance can be written as follows:
\begin{equation}
R_{2\omega,{\rm H}}=~R_{\rm FL+Oe}~(2{\rm cos^3}~{\varphi}~-~{\rm cos}~{\varphi})-\frac{1}{2}(R_{\rm DL}+R_{\rm ANE})~{\rm cos}~{\varphi}
\label{eq:two}.
\end{equation}
The different $\varphi$-dependence of ${R_{\rm FL+Oe}}$ and ${R_{\rm DL}}$ make the $xy$ angle scan the most convenient measurement method to separate the two contributions. Further, these SOT resistances are inversely proportional to the external magnetic field, the increase of which progressively suppresses the torques' action. $R_{\rm ANE}$ shares the same $\varphi$-dependency with $R_{\rm DL}$ but it is field-independent. Thus, the two terms can be ultimately separated via their different magnetic field-dependence. For these reasons, we quantified the SOTs by performing $xy$ angle scan measurements at different external magnetic fields $B_{\rm ext}$.

Figure~\ref{fig:FIGURE3}(a) shows the 1$\textsuperscript{st}$ harmonic Hall signal of the sample $\alpha$-Sn/Bi for $xy$ angle scans at three representative fields. The main signal contribution arises from the planar Hall effect, whose corresponding resistance $R_{\rm PHE}$ increases with $B_{\rm ext}$, indicating that the magnetization is not completely saturated for the smaller applied fields. Figure~\ref{fig:FIGURE3}(b) shows the $2\textsuperscript{nd}$ harmonic Hall signal recorded simultaneously with the 1$\textsuperscript{st}$ harmonic signal. We fitted $R_{2\omega,{\rm H}}$ using Eq.~\ref{eq:two}, finding the two coefficients $(R_{\rm FL+Oe})$ and $(R_{\rm DL}+R_{\rm ANE})$ at different $B_{\rm ext}$.

We quantified the FL-SOT by exploiting the field-dependent relation $R_{\rm FL+Oe}= \frac{1}{2}~\frac{R_{\rm PHE}}{B_{\rm eff}}~{B_{\rm FL+Oe}}$, with ${B_{\rm eff}=B_{\rm ext}}$ \cite{G50}. Figure~\ref{fig:FIGURE3}(c) presents $R_{\rm FL+Oe}$ versus $\frac{1}{B_{\rm ext}}$ (red dots), with each data point rescaled by a factor $\frac{R_{\rm PHE}}{{\rm max}(R_{\rm PHE})}$ to account for the varying $R$\textsubscript{PHE} at different $B$\textsubscript{ext}, with ${\rm max}(R_{\rm PHE})$ being the maximum value of $R$\textsubscript{PHE}. In this type of plot, $B\textsubscript{FL+Oe}$ corresponds to the slope of the fit (red dashed line) divided by $\frac{{\rm max}(R_{\rm PHE})}{2}$. \color{black} Next, to determine ${B_{\rm FL}}$, we subtracted the calculated Oersted field $B_{\rm Oe}$ from ${B_{\rm FL+Oe}}$, taking $B_{\rm Oe}=\frac{\mu_0 {I_{\rm Sn}}}{2 w}$, where $\mu_0$ is the vacuum permeability, $w$ the current line width and ${I_{\rm Sn}}$ the current flowing in the nonmagnetic layer. ${I_{\rm Sn}}$ was estimated using a parallel resistor model for the Sn/FeCo bilayer by measuring the four-point resistance of the full heterostructures and single FeCo films grown on CdTe and MgO substrates.  Table~\ref{tab:table00} presents the resistivity and the four-point resistance of the full heterostructures ($\rho$, $R$) and single FeCo films ($\rho_{\rm FeCo}$, $R_{\rm FeCo}$) together with the resultant branching of the current in the Sn and FeCo layers, defined as $\frac{I_{\rm Sn}}{I}$ and $\frac{I_{\rm FeCo}}{I}$, where $I$ is the total current sent through the full heterostructure.

\begin{table*}[h!]
 \color{black}
\centering
\begin{adjustbox}{max width=\textwidth}
\small
\begin{tabular}{llllllllll}
\hline \hline
Sample & $\rho$ & $R$ & $\rho_{\rm FeCo}$ & $R_{\rm FeCo}$ & $\frac{I_{\rm Sn}}{I}$ & $\frac{I_{\rm FeCo}}{I}$ & $I$ & $j$ & $j\textsubscript{Sn}$\\

 & [$\mu\Omega$cm] & [k$\Omega$] & [$\mu\Omega$cm] & [k$\Omega$]  & [\%] & [\%] & [mA] & [10$\textsuperscript{6}$~A/cm$\textsuperscript{2}$] & [10$\textsuperscript{6}$~A/cm$\textsuperscript{2}$] \\
 
\hline
$\alpha$-Sn/Bi & 179 & 1.3 & 72  & 2.1 & 42  & 58 & 2 & 2.84  & 1.68 \\
$\alpha$-Sn    & 204 & 1.5 & 72  & 2.1 & 31  & 69 & 2 & 2.98  & 1.24 \\
$\beta$-Sn     & 342 & 2.4 & 224 & 6.6 & 63  & 37 & 2 & 2.98  & 1.18 \\
\hline \hline
\end{tabular}
\end{adjustbox}

\caption{\label{tab:table00} \color{black} Summary of the electrical properties of $\alpha$-Sn/Bi, $\alpha$-Sn, and $\beta$-Sn obtained using a parallel resistor model at room temperature. From left to right: resistivity and four-point resistance  of the full heterostructure and single FeCo film grown on different substrates, current branching ratio in the Sn  and FeCo layer, current and  average current density flowing through the full heterostructure, current density in the Sn layer.}
 
\end{table*}

 \color{black}

We quantified the DL-SOT by considering the different field dependence of $B_{\rm DL}$ and the magnetothermal effects due to the out-of-plane thermal gradient produced by Joule heating. More specifically, we accounted for the anomalous Nernst effect represented by the resistance ${R_{\rm ANE}}$. This effect depends exclusively on the magnetization and thermal gradient, and is thus independent of $B_{\rm ext}$. To separate ${R_{\rm ANE}}$ from ${R_{\rm DL}}$ in Eq.~\ref{eq:two}, we used the relation  $({R_{\rm DL}+{R_{\rm ANE}})= -\frac{1}{2}~\frac{R_{\rm AHE}}{B_{\rm eff}}~{B_{\rm DL}}+{R_{\rm ANE}}}$, where ${B_{\rm eff}=B_{\rm ext}+B_{\rm dem}+B_{\rm PMA}}$ \cite{G50}.  Figure.~\ref{fig:FIGURE3}(c) shows $({R_{\rm DL}+R_{\rm ANE}})$ versus ${1/({B_{\rm ext}+B_{\rm dem}+B_{\rm PMA}})}$ (black dots). In this plot, $B\textsubscript{DL}$ corresponds to the slope of the fit (black dashed line) divided by $-\frac{R_{\rm AHE}}{2}$, whereas ${R_{\rm ANE}}$ corresponds to the intercept with the y-axis, the only residual resistance contribution when ${B_{\rm ext}\rightarrow\infty}$.

For the $\alpha$-Sn/Bi sample, we additionally took into account the ordinary Nernst effect (ONE), given by $R_{\rm ONE}B_{\rm ext}{\rm cos}~{\varphi_{B}}$, where $\varphi_{B}$ is the azimuthal angle of the applied field $B_{\rm ext}$.  This term adds to Eq.~\ref{eq:two} a resistance proportional to $B_{\rm ext}$ and the out-of-plane thermal gradient. Spintronic devices containing Bi and Bi-based alloys often exhibit a large ONE due to their low thermal conductivity and large Nernst coefficient \cite{yue2018spin}. In these systems, the ONE can play an important role for properly quantifying the SOTs \cite{roschewsky2019spin}. For $\alpha$-Sn/Bi, modifications of the electronic band structure or growth quality induced by Bi \cite{G27} could also affect the thermoelectric properties of the stack. Experimentally, we measured $R_{\rm ONE}B_{\rm ext}$ by performing a field scan at $\varphi_{B}$=0 with $B_{\rm ext}>1.5~{\rm T}$, where $R_{\rm FL+Oe}$ and $R_{\rm DL}$ are suppressed and $R_{\rm ANE}$ is constant. For $j$=2.8$\times$10\textsuperscript{-6}~A/cm\textsuperscript{2} and $B_{\rm ext}=1.5~$T, we found $R_{\rm ONE}B_{\rm ext}$=1.8$\times$10\textsuperscript{-4}~$\Omega$, which is about half of $R_{\rm ANE}$. We then subtracted the measured $R_{\rm ONE}B_{\rm ext}~{\rm cos}~{\varphi_{B}}$ to the raw 2$\textsuperscript{nd}$ harmonic Hall resistance signal of $\alpha$-Sn/Bi. In contrast, in $\alpha$-Sn and $\beta$-Sn, we found a negligible ONE.  

 Figure~\ref{fig:FIGURE3}(d) shows ${B_{\rm DL}}$ and ${B_{\rm FL}}$ after subtraction of ${B_{\rm Oe}}$ as a function of $j$, together with the estimated ${B_{\rm Oe}}$ (dashed line) in $\alpha$-Sn/Bi. The red and black lines are fits forced to intercept at zero, which serve as guides to the eye. The good agreement between fits and data proves the linear relation between SOTs and injected current density. By using the same method, we quantified the SOT effective fields in the $\alpha$-Sn and $\beta$-Sn samples. In the next section we compare the SOT efficiency for all samples.

\subsection{Effective spin-orbit fields and SOT efficiencies}

\begin{figure}[h]
\includegraphics[scale=0.46]{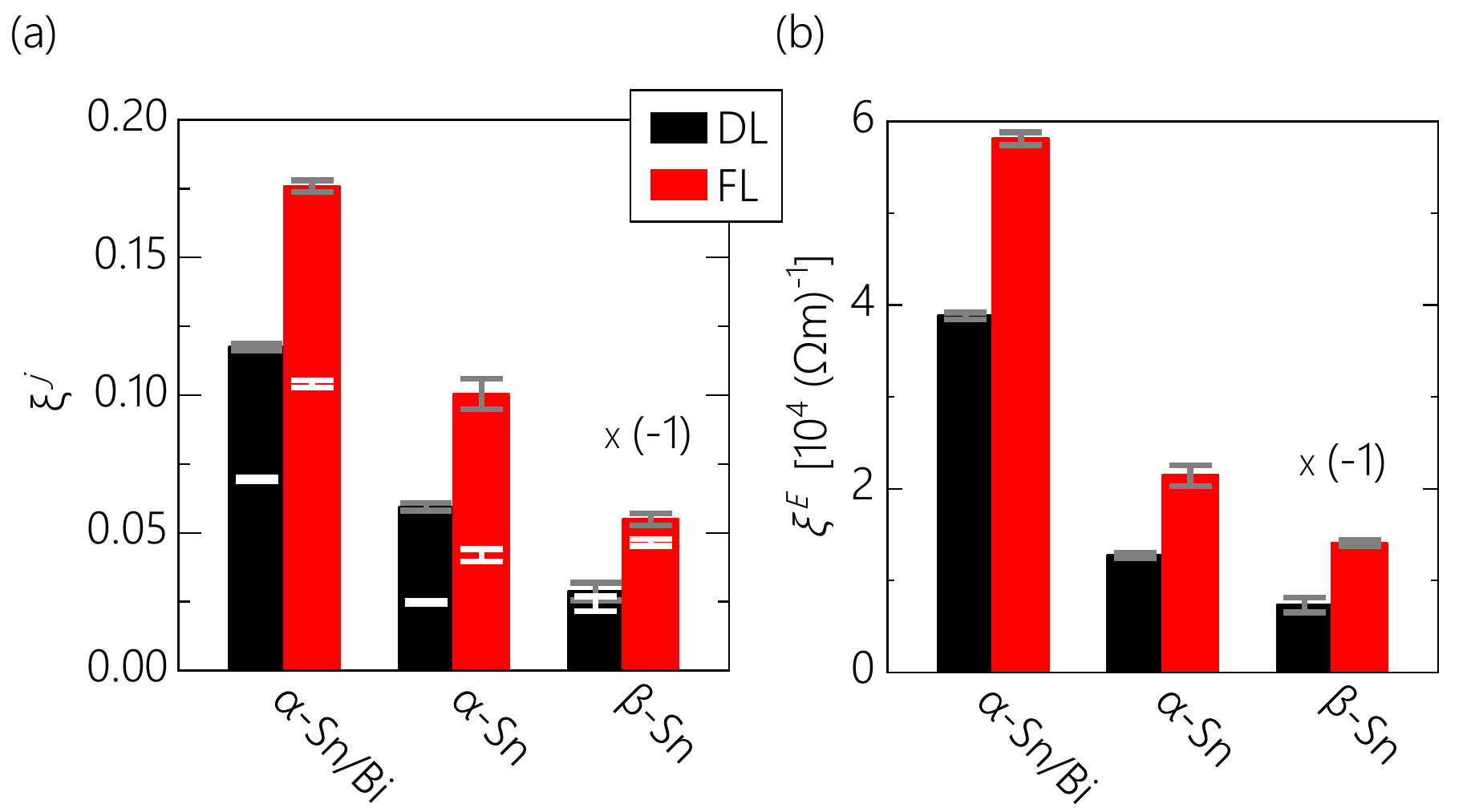}
\caption{\label{fig:FIGURE4} Comparison of the SOT efficiency in $\alpha$-Sn/Bi, $\alpha$-Sn and $\beta$-Sn. (a) Efficiency with respect to the current density, showing $\xi_{\rm DL,FL}^{j_{\rm Sn}}$ as colored bars and $\xi_{\rm DL,FL}^{j}$ as white error bars. $j\textsubscript{Sn}$ refers to the current density considering the partial current exclusively flowing through the Sn layer, whereas $j$ refers to the average current density flowing through the full thickness of the heterostructure. (b) Efficiency with respect to the electric field. The efficiencies of $\beta$-Sn are multiplied by -1 to account for the inverted structure.}
\end{figure}

The SOT effective fields measured in different samples can be compared by converting them into so-called SOT efficiencies \cite{G19, G117, G217}. There are two ways to do so. The first one is by normalizing the SOT effective fields to the current density, and the second one is by normalizing then to the applied the electric field, which depends on the resistivity but is independent of the current distribution in the heterostructure.  The two efficiencies are respectively defined as follows:
\begin{equation}
\xi_{\rm DL,FL}^{j}=~\frac{2e}{\hbar}~{M_s}~{t_F}~\frac{B_{\rm DL,FL}}{j},
\label{eq:three}
\end{equation}
\begin{equation}
\xi_{\rm DL,FL}^{E}=~\frac{2e}{\hbar}~{M_s}~{t_F}~\frac{B_{\rm DL,FL}}{E},
\label{eq:four}
\end{equation}
where $e$ is the elementary charge, $\hbar$ the reduced Planck constant, $j$ the current density, $E$ the applied electric field, and  $t_F$ the thickness of the ferromagnet. 
$\xi_{\rm DL,FL}^{j}$ represents the ratio of the spin current absorbed by the ferromagnet relative to the injected charge current, which can be considered as an effective spin Hall angle. $\xi_{\rm DL,FL}^{E}$, on the other hand, represents the ratio of the spin current absorbed by the ferromagnet relative to the applied electric field, that is, the effective spin Hall conductivity. Considering that spin reflection and memory loss take place at the interface between nonmagnetic and magnetic materials, the efficiencies calculated by the above formulae constitute the lower bound for quantifying the charge-to-spin conversion capacity of a nonmagnetic layer.

Figure~\ref{fig:FIGURE4} (a) and (b) compares the efficiencies of DL-SOT (black columns) and FL-SOT (red columns, after subtraction of the  Oersted field contribution) for the three samples with respect to $j$ and $E$, respectively.  \color{black}  We note that the definition of the current density in the context of SOT efficiency may be ambiguous. Some consider this term as the average current density flowing through the full thickness of the heterostructure (here defined as $j$), whereas others apply a parallel resistor model to separate the currents flowing through the nonmagnetic metal (here defined as $j_{\rm Sn}$) and ferromagnetic layer. The estimated values of $j$ and $j_{\rm Sn}$ are shown in Table~\ref{tab:table00}. \color{black} In Fig.~\ref{fig:FIGURE4}(a) we account for both current densities, showing $\xi_{\rm DL,FL}^{j_{\rm Sn}}$ as colored bars and $\xi_{\rm DL,FL}^{j}$ as white error bars. Depending on the resistance ratio between the nonmagnetic and ferromagnetic layers, $\xi_{\rm DL,FL}^{j_{\rm Sn}}$ and $\xi_{\rm DL,FL}^{j}$ can differ significantly from each other, as seen by comparing $\alpha$-Sn and $\beta$-Sn in Fig.~\ref{fig:FIGURE4}. In any case, we find that $\alpha$-Sn/Bi shows the largest DL- and FL-SOT efficiency, followed by $\alpha$-Sn and $\beta$-Sn.

\begin{table*}
\centering
\begin{adjustbox}{max width=\textwidth}
\small
\begin{tabular}{lllllllllllll}
\hline \hline
Sample &$\rho$ &$M_{\rm s}$ &$B_{\rm sat}$ &$K_{\rm \perp}$ &$R_{\mathrm{AHE}}$ &$\frac{\Delta R_{\rm xy}}{R_{0}}$&$\frac{\Delta R_{\rm zy}}{R_{0}}$&$\frac{\Delta R_{\rm zx}}{R_{0}}$ &$\xi_{\rm DL}^{\textit{$j_{\rm Sn}$}}$  &$\xi_{\rm FL}^{\textit{$j_{\rm Sn}$}}$ &$\xi_{\rm DL}^{\textit{E}}$ &$\xi_{\rm FL}^{\textit{E}}$  \\
 
 &[$\mu\Omega$cm] &[10$\textsuperscript{6}$\rm{A/m]} &[mT] &[10$\textsuperscript{5}\rm{J/m\textsuperscript{3}]}$ &[$\Omega$] &[$\%$]& [$\%$]&[$\%$] &$\times10\textsuperscript{-2}$ &$\times10\textsuperscript{-2}$ &[10\textsuperscript{4}~($\Omega$m)\textsuperscript{-1}] &[10\textsuperscript{4}~($\Omega$m)\textsuperscript{-1}]\\
\hline
$\alpha$-Sn/Bi  &179   &0.7 &470  &1.4 &1.1 &0.02 &0.02 &$\approx$~0       &12.0~$\pm$~0.1    &17.6~$\pm$~0.1   &3.9~$\pm$~0.1    &5.8~$\pm$~0.1\\
$\alpha$-Sn     &204   &1.0   &625  &3.2 &3.5 &0.05 &0.05 &$\approx$~0       &5.9~$\pm$~0.1    &10.0~$\pm$~0.2    &1.3~$\pm$~0.1    &2.1~$\pm$~0.1\\
$\beta$-Sn      &342   &1.5 &780  &8.3 &5.4 &0.05 &0.015 &-0.035 &-2.9~$\pm$~0.3   &-5.5~$\pm$~0.1   &-0.7~$\pm$~0.1   &-1.4~$\pm$~0.1\\
\hline \hline
\end{tabular}
\end{adjustbox}

\caption{\label{tab:table0} Summary of the magnetic properties and SOT efficiencies of $\alpha$-Sn/Bi, $\alpha$-Sn, and $\beta$-Sn. All data measured at room temperature. From left to right: resistivity, saturation magnetization, out-of-plane saturation field, perpendicular magnetic anisotropy constant, anomalous Hall resistance, normalized magnetoresistance for xy, zy and zx angular scans, DL- and FL-SOT efficiency normalized to the current density and to the electric field.}
\end{table*}

\section{Discussion}

In this section, we compare our findings with the literature and examine the correlation between the SOT efficiency and magnetotransport properties of our samples, and discuss the role of the Bi surfactant. The most relevant properties on which we base our discussion are summarized in Table~\ref{tab:table0}. 

\subsection{Comparison of SOTs with literature}

Here we compare the SOTs efficiency of the $\alpha$-Sn samples with the literature, and particularly with the spin-pumping measurements of the only previously studied $\alpha$-Sn-based ferromagnetic system: $\alpha$-Sn[50~\AA]/Bi[3.5~\AA]/Ag[0~and~20~\AA]/Fe~[50~\AA] \cite{G31}. The Ag spacer layer in this study was used to preserve the topological insulator nature of the $\alpha$-Sn film. A monolayer of Bi was used also in this system to improve the growth of $\alpha$-Sn \cite{G27, G31}. With the Ag spacer, the spin-to-charge conversion efficiency was reported to be 0.62, a value one order of magnitude larger than Pt, considering a spin diffusion length of 3.4~nm. Without spacer, the conversion efficiency was shown to be negligible, in line with the disappearance of the Dirac cone observed by angle-resolved photoemission \cite{G31}. Interestingly though, we measure a sizeable SOT efficiency (or charge-to-spin conversion) in $\alpha$-Sn samples in direct contact with the ferromagnet. Our values of $\xi_{\rm DL}^{j_{\rm Sn}}$=0.12 (0.06), $\xi_{\rm FL}^{j_{\rm Sn}}$=0.18 (0.10) in $\alpha$-Sn with Bi (without Bi) are lower but still significant compared the spin-to-charge conversion efficiency measured by spin pumping in Ref.~\onlinecite{G31}. In making such a comparison, one should note that the conversion efficiency of the same material, measured by different techniques and on different samples can vary. In particular, TIs exhibit $\xi_{\rm DL}^{j}$ spreading over several orders of magnitude from below 0.01 \cite{deorani2014observation} to more than 1 \cite{G123}. In our case, we cannot exclude that the use of a different substrate or measurement geometry could affect the conversion efficiency. Nevertheless, our results support the finding that $\alpha$-Sn is an efficient material for spin-charge interconversion. Additionally we evidenced that the FL-torque is large and even larger than the DL-torque; a property that could not be studied by spin pumping.

Next, we compare the SOT efficiencies obtained in $\alpha$-Sn/Bi with other SOT materials. Pt is by far the most studied single-element metal in the context of SOTs, with reported efficiencies per current density in the range of $\xi_{\rm DL,~Pt}^{j}$=0.06-0.2,~$\xi_{\rm FL,Pt}^{j}$=-0.01-0.12 \cite{G117,G19}. Our values for $\xi_{\rm DL}^{j}$ are comparable to Pt. The efficiencies per electric field, $\xi_{\rm DL}^{E}$=0.39$\times10^{5}$~($\Omega$m)\textsuperscript{-1} and $\xi_{\rm FL}^{E}$=0.58$\times10^{5}$~($\Omega$m)\textsuperscript{-1}, are lower by a factor of 5-10 compared to the spin Hall conductivity of Pt because $\alpha$-Sn has a significantly larger resistivity than Pt. In terms of torque direction, we find that the SOTs generated by $\alpha$-Sn have the same sign as in Pt/Co with the same stacking order between nonmagnetic and magnetic layers \cite{G19}. The SOTs efficiencies of $\alpha$-Sn/Bi are compatible with the lower range of the efficiencies found for the Bi-chalcogenide TIs, which are spread out over two orders of magnitude. For example, BiSe\textsubscript{3} has $\xi_{\rm DL/FL}^{j}=$0.3-18 and $\xi_{\rm DL/FL}^{E}$=1-1.5$\times10^{5}$~($\Omega$m)\textsuperscript{-1}) \cite{G13, G123, G130, G132, G211}, $\rm{(Bi,Sb)}_{2}\rm{Te}_{3}$ has $\xi_{\rm DL/FL}^{j}=$0.4-3 \cite{G129, wu2019spin, G211}, and Bi\textsubscript{x}Sb\textsubscript{1-x} has $\xi_{\rm DL/FL}^{j}=$0.5-52 \cite{G126, G197}. These values fluctuate significantly depending on TI composition, current normalization and TI thickness, as well as on the measurement method and conditions, which hinders a stricter comparison of our results with other TI systems. Interestingly, our measurements also show that the non-topological phase $\beta$-Sn presents finite DL- and FL-SOT efficiency, which are, however, about a factor four smaller relative to $\alpha$-Sn/Bi. 

\subsection{Role of the Bi surfactant}

\color{black}

Since the SOT efficiencies are significantly higher in $\alpha$-Sn/Bi compared to $\alpha$-Sn, we further comment on the role of the Bi surfactant. A previous core-level photoemission study has shown that the Bi atoms segregate on top of epitaxial $\alpha$-Sn grown on InSb(001)  \cite{G27}. Our angle-dependent XPS data qualitatively show that Bi locates near the surface of our $\alpha$-Sn-based heterostructure, despite being the first layer to be deposited on the CdTe substrate. This indicates that Bi has a strong tendency to diffuse through the deposited layers, behaving as a surfactant for Sn grown on CdTe and also segregating on top of the magnetic layer. An indication that the Bi atoms are located in contact with the FeCo layer additionally comes from the lower $R_{\rm AHE}$ of $\alpha$-Sn/Bi with respect to $\alpha$-Sn, despite their comparable resistivity (see Table~\ref{tab:table0}). Sagasta \textit{et al.} showed a similar reduction of $R_{\rm AHE}$ when a bismuth oxide layer is placed next to Co, attributing it to an interfacial skew scattering contribution to the Hall resistance of opposite sign with respect to the bulk AHE in Co \cite{G189}.

Considering the high diffusivity of Bi atoms, and depending on their exact location, Bi may play different roles in generating the SOTs. In the case of Bi atoms locating between the FeCo and the AlO\textsubscript{x} layers, the interfacial region Bi/FeCo may become a source of SOTs itself due to the non-topological Rashba-Edelstein effect, as it is the case for Bi/Ag surface alloy \cite{sanchez2013spin, G138}. These SOTs, generated in the upper side of FeCo, would then add to the SOTs generated by the $\alpha$-Sn layer below, eventually leading to larger SOT in the $\alpha$-Sn/Bi sample. Another possibility is that $\alpha$-Sn films grown with Bi as a surfactant have enhanced interfacial topological properties relative to $\alpha$-Sn grown without Bi. In the first angle-resolved photoemission study of $\alpha$-Sn grown on InSb(001), it was shown that $\alpha$-Sn covered by a monolayer of Bi presents topological surface states that have a sharper and more isotropic dispersion relative to the bare $\alpha$-Sn surface \cite{G27}. Therefore, together with its surfactant effect, Bi might promote the formation of spin-momentum locked topological surface states that contribute to the charge-to-spin conversion. Finally, some Bi atoms might be incorporated in the FeCo and Sn layers, which would also affect the generation and absorption of a spin current. 

The exact role of Bi thus remains ambiguous, in particular whether it leads to a significant improvement of the topological properties of the interface or it is a source of interfacial Rashba-Edelstein effect. Regarding the ultimate enhancement of charge-to-spin conversion given by Bi in $\alpha$-Sn-based system, we note  that also the spin pumping results in Ref.~\onlinecite{G31} have been obtained on $\alpha$-Sn/Ag/Fe grown with a Bi surfactant, which might enhance the spin-to-charge conversion efficiency when in contact with Ag \cite{sanchez2013spin}.

\color{black}

\subsection{Correlation between magnetotransport properties and SOTs}

In Table~\ref{tab:table0}, we observe a direct correlation between the perpendicular magnetic anisotropy (PMA), represented by the constant $K_{\rm \perp}$ and the anomalous Hall effect (AHE), represented by the Hall resistance $R_{\rm AHE}$. Both decrease going from $\beta$-Sn to $\alpha$-Sn and finally $\alpha$-Sn/Bi. Such a decrease is too large to be ascribed solely to the reduction of $M_s$ in these samples. Moreover, $K_{\rm \perp}$ and $R_{\rm AHE}$ are inversely correlated with the SOT efficiencies, i.e., the stronger the SOTs are, the lower the PMA and AHE are. 
This inverse correlation is in striking contrast with the direct correlation between $K_{\rm \perp}$ and the DL-SOT observed in heterostructures where 5$d$ metals are the main SOT source \cite{G182, pai2014enhancement, xie2019giant} or where a transition-metal spacer is inserted between Pt or W and the ferromagnetic layer \cite{G182, pai2014enhancement}. In general, the PMA and AHE are very sensitive to the spin-orbit coupling at the interface. If we compare $\alpha$-Sn/Bi and $\alpha$-Sn, for which the only difference is the use of the Bi surfactant, it is clear that Bi significantly decreases the PMA and AHE of FeCo while enhancing the SOT efficiency. \color{black} This behavior might be tentatively explained by assuming that Bi induces an interfacial skew scattering contribution to the AHE of opposite sign with respect to the bulk AHE in FeCo, similar to Co/Bi$\textsubscript{2}$O$\textsubscript{3}$ bilayers \cite{G189}, and enhances at the same time the generation of SOTs. The SOTs might then be increased by the generation of an additional Rashba-interface on top of FeCo, or by promoting the formation of spin-momentum locked interface states, or eventually by the formation of a spin current by skew scattering \cite{sanchez2013spin, G138, G27, cho2015large, honemann2019spin}. \color{black} An enhancement of the SOTs due to intermixing of Bi and Sn in the bulk of the $\alpha$-Sn layer appears less likely given the tendency of Bi to segregate on top of Sn \cite{G27} and the strong reduction of the PMA and AHE in $\alpha$-Sn/Bi relative to $\alpha$-Sn. 

Another unusual inverse correlation is that between the SOT efficiency and the SMR-like magnetoresistance in $\alpha$-Sn/Bi and $\alpha$-Sn (Table~\ref{tab:table0}). Measurements of the SMR and SOTs in transition-metal bilayers reveal a direct correlation between these two quantities \cite{G182, cho2015large, kim2016spin}. Instead, we observe that $\frac{\Delta R_{\rm xy}}{R_{0}}\approx\frac{\Delta R_{\rm zy}}{R_{0}}\approx~$0.05 \% is larger in $\alpha$-Sn compared to $\alpha$-Sn/Bi, where $\frac{\Delta R_{\rm xy}}{R_{0}}\approx\frac{\Delta R_{\rm zy}}{R_{0}}\approx~$0.02 \%, while the SOT efficiency is larger in $\alpha$-Sn/Bi. This might be understood considering the different dependencies that SMR and SOT efficiency have on magnetization and film thickness. The SMR amplitude can be reduced when thickness of the magnetic layer is too small \cite{karwacki2020optimization} possibly due to a reduced magnetization \cite{yang2016thickness}. The $M$\textsubscript{s} of $\alpha$-Sn/Bi is smaller than the one of $\alpha$-Sn despite the similar thickness of the ferromagnetic layer, likely affecting the SMR amplitude. Additionally, as seen in the drift-diffusion model of the spin-Hall effect with spin-Hall angle ${\theta_{\rm SHE}}$, we note that the SOT efficiency and the SMR do not have the same dependency upon the spin diffusion length $\lambda_{s}$. In the spin Hall picture, the SMR amplitude is proportional to ${\theta_{\rm SHE}}^{2}\lambda_{s}$, whereas the SOT efficiency is directly proportional to ${\theta_{\rm SHE}}$ \cite{chen2013theory}. Therefore a reduction of $\lambda_{s}$ due to the addition of Bi could lead to an enhanced torque efficiency but a reduced SMR signal. The magnetoresistance behavior of $\beta$-Sn, on the other hand, is more similar to that expected for the anisotropic magnetoresistance of FeCo \cite{li2019giant}, in agreement with the smaller charge-to-spin conversion derived from the SOT measurements.

\section{Conclusions}

In summary, we have characterized the epitaxial growth of Sn on CdTe(001) and Mg/FeCo(001) showing that single-phase $\alpha$-Sn and $\beta$-Sn epitaxial films can be obtained by proper choice of the substrate. All the FeCo films, which are 17~\AA -thick, present in-plane magnetization driven by shape anisotropy and a competing out-of-plane magnetic anisotropy that is largest for $\beta$-Sn where FeCo is grown on MgO. Our measurements evidence the emergence of SMR-like magnetoresistance in FeCo grown on $\alpha$-Sn, which is a signature of current-induced spin accumulation in this system. The dampinglike and fieldlike SOT efficiencies are larger in $\alpha$-Sn/Bi ($\xi_{\rm DL}^{j_{\rm Sn}}$=0.12, $\xi_{\rm FL}^{j_{\rm Sn}}$=0.18), intermediate in $\alpha$-Sn ($\xi_{\rm DL}^{j_{\rm Sn}}$=0.06, $\xi_{\rm FL}^{j_{\rm Sn}}$=0.10), and smaller in $\beta$-Sn ($\xi_{\rm DL}^{j_{\rm Sn}}$=0.03, $\xi_{\rm FL}^{j_{\rm Sn}}$=0.06). We further found an inverse correlation between the SOT efficiency and the AHE and PMA, highlighting the importance of the Bi surfactant for improving the SOT efficiency and  modifying  the  magnetotransport  properties in $\alpha$-Sn-based ferromagnetic heterostructures. Our results show that $\alpha$-Sn grown with a Bi surfactant is an efficient material for direct charge-to-spin conversion and generation of SOTs.

\section{Acknowledgements}
This work was supported by the Swiss National Science Foundation (Grant No. 200020-172775). Paul N\"{o}el acknowledges support from the ETH Postdoctoral Fellowship program. The authors thank Yoshiyuki Ohtsubo for fruitful discussions about the growth of $\alpha$-Sn. 


%

\clearpage

\end{document}